\documentclass[1p]{elsarticle}
\usepackage{epsfig}
\usepackage{epsf}

\def\slash#1{\setbox0=\hbox{$#1$}
   \dimen0=\wd0 \setbox1=\hbox{/} \dimen1=\wd1
   \ifdim\dimen0>\dimen1 \rlap{\hbox to \dimen0{\hfil/\hfil}} #1
   \else  \rlap{\hbox to \dimen1{\hfil$#1$\hfil}} / \fi}
   
\begin{document}

\title{Correlations equalities and some upper bounds for the critical temperature for spin one systems. }

\author[rvt]{F. C. S\'{a} Barreto}
\ead{fcsabarreto@gmail.com}
\author[rvt]{A. L. Mota}
\ead{motaal@ufsj.edu.br}
\address[rvt]{Departamento de Ci\^{e}ncias Naturais, Universidade Federal de S\~{a}o Jo\~{a}o del Rei, C.P. 110,  CEP 36301-160, S\~ao Jo\~ao del Rei, Brazil}

\begin{abstract}
Starting from correlation identities for the Blume-Capel spin 1 systems and using correlation inequalities, we obtain rigorous upper bounds for the critical temperature.The obtained results improve over effective field type results. 
\end{abstract}

\maketitle

\section{Introduction}
Correlation inequalities combined with exact identities are useful in obtaining
rigorous results in statistical mechanics. Among the various questions that are
resolved by them one is the decay of the correlation functions.
The decay of the correlation functions give information about the critical
couplings of statistical mechanics models. In this work, the method will
be applied to study systems described by the spin one Blume-Capel model \cite{Capel1966,Blume1966}.
Firstly, we present the derivation of an exact relation for the two spin correlation
function, valid in any dimension, which is an extension of Callen's identity
for spin $1/2$ Ising model \cite{Callen1963}. Starting from these identities we will then make use of
the first and second Griffiths inequalities and Newman's inequalities to obtain
the exponential decay of the two spin correlation function . The coupling constant which
are the upper bounds for the critical temperature are obtained for d=2 and
d=3 dimensions. In this study the coupling parameters obtained improve effective field results.
Upper bounds for the critical temperature $T_c$ for Ising and multi-component spin systems have been obtained by showing (for $T>T_c$) the exponential decay of the two-point function \cite{Fisher1967, Simon1980, Brydges1982}. Spin correlation inequalities and their iteration are used by Brydges et al \cite{Brydges1982}, Lieb \cite{Lieb1980} and Simon \cite{Simon1980}. 
The procedure to improve the bound for the critical temperature over the effective field result for the classical $S=1$ model is as follows: starting from a two-point correlation function identity, a generalization of Callen's identity \cite{Callen1963} for this model \cite{SiqueiraandFittipaldi1986} and using Griffith�s 1st and 2nd inequalities (Griffith I, II) (see \cite{Griffiths1969}, \cite{Sylvester1976},\cite{Fernandez},\cite{Braga1},\cite{Braga2},\cite{Szasz1978}) and Newman's inequalities \cite{Sylvester1976,Newman1975} we establish the inequality for the two-point function, $<S_0S_l>$, as
\begin{equation}
<S_0 S_l> \leq  \sum_j a_j <S_j S_l>, 0\leq a_j\leq 1 
\end{equation}
which when iterated (see \cite{Simon1980}) implies exponential decay for $T>T_c$. 
In section 2 we present the derivation of the correlation identities for the Blume-Capel model \cite{SiqueiraandFittipaldi1986}. In section 3, we apply these identities to the $d=2$ and $d=3$ lattices. Next, in section 4, we apply the correlation inequalities to obtain the upper bounds for $T_c$. Numerical results can be found in section 5, and in section 6 we present our concluding remarks.

We write the Hamiltonian for the classical spin one system, known as the Blume-Capel model, as
\begin{equation}
H = -J \sum_{i,j} S_i S_j - D \sum_i S_i^2,  \label{eq2}
\end{equation}
where $J>0$, $D$ is the single ion anisotropy and the first sum is over the nearest neighbours spins on the lattice. 
We define the thermal average $< . . .>$ by
\begin{equation}
 <...>=Z^{-1} \sum_{\{S_i\}} (...) e^{-\beta H},  Z=\sum_{\{S_i\}} e^{-\beta H} \label{eq3}
\end{equation}
where each $S_i$ is restricted by $S_i = -1,0,+1$. 


 

\section{Correlation identity for the spin one model}

We reproduce the generalization of Callen's identity for the spin 1 Blume-Capel model which has been obtained previously by Siqueira and Fittipaldi \cite{SiqueiraandFittipaldi1986}. Let 
\begin{equation}
<F{(S)} S_i> = \frac{ Tr(F{(S)} S_i e^{-\beta H})}{Tr(e^{-\beta H})}, 
\end{equation}
where $F(S)$ is any function of $S$ different from $S_i$.
We can write $H = H_i + H^\prime$, where
\begin{equation}
H_i = - (\sum_{|j|=1} J_{ij}S_j)S_i - D S_i^2, 
\end{equation}
is the Hamiltonian describing site i and its neighbours, 
and $H^\prime$ corresponds to the Hamiltonian of the rest of the lattice. Consequently $[H_i,H^\prime]=0$.
From Eq.(\ref{eq2}) and Eq.(\ref{eq3}), we get,
\begin{equation}
<F{(S)}S_i>=\frac{Tr F{(S)}e^{-\beta (H_i + H')} S_i}{Tr e^{-\beta (H_i + H')}} =\frac{Tr' Tr_i F{(S)}e^{-\beta H_i} S_i e^{-\beta H'}}{Tr' Tr_i e^{-\beta H_i} e^{-\beta H'}}
\end{equation}
or
\begin{equation}
<F{(S)}S_i>=\frac{Tr' Tr_i F{(S)}e^{-\beta H_i} e^{-\beta H'} \frac{Tr_i e^{-\beta H_i} S_i}{Tr_i e^{-\beta H_i}}}{Tr' Tr_i e^{-\beta H_i} e^{-\beta H'}}
\end{equation}
where $Tr' Tr_i = Tr$. Finally, we obtain,
\begin{equation}
<F{(S)}S_i>= \Big<F{(S)} \frac{Tr_i e^{-\beta H_i} S_i}{Tr_i e^{-\beta H_i}} \Big>. 
\end{equation}

Explicitly operating the trace $Tr_i$, we get,
\begin{eqnarray}
<F{(S)}S_i>&=&\Big<F{(S)} \frac{2e^{\beta D} sinh(\sum_j \beta J_{ij} S_j)}{2e^{\beta D} cosh(\sum_j \beta J_{ij} S_j)+1} \Big> \nonumber \\
&=&\Big< F{(S)}\prod_{|j|=1} e^{\beta J_{ij} S_j \nabla} \Big> f(x)|_{x=0} , \label{eq9}
\end{eqnarray}
with $\nabla$ $\equiv$  $\frac{\partial}{\partial x}$, such that $e^{\alpha\nabla} f(x) = f(x+\alpha)$, and
\begin{equation}
f(x)=\frac{2 e^{\beta D} sinh(x)}{2 e^{\beta D} cosh(x) + 1}. \label{eq10}
\end{equation}
As $S_j^{2n}=S_j^2$ and $S_j^{2n+1}=S_j$ for $n=0,1,2,3,...$, we obtain,
\begin{eqnarray}
e^{S_j A}&=& S_j^2 cosh(A) + S_j sinh(A) + 1 - S_j^2, \label{eq11}
\end{eqnarray}
and, applying Eq.(\ref{eq10}) and Eq.(\ref{eq11}) in Eq.(\ref{eq9}), we get
\begin{equation}
<F{(S)}S_i>= \Big<F{(S)} \prod_{j \neq i,|j|=1} (S_j^2 cosh(\beta J_{ij} \nabla ) + S_j sinh(\beta J_{ij} \nabla ) + 1 - S_j^2 ) \Big> f(x)|_{x=0} \label{eq12}
\end{equation}
Similarly for the correlation function involving the square of the spin  function $S_i^2$, we obtain,
\begin{equation}
<G{(S)}S_i^2> = \Big<G{(S)} \prod_{j \neq i,|j|=1} e^{\beta J_{ij}S_j \nabla } \Big> g(x)|_{x=0},
\end{equation}
with,
\begin{equation}
g(x)=\frac{2 e^{\beta D} cosh(x)}{2 e^{\beta D} cosh(x) + 1}, \label{eq14}
\end{equation}
resulting in,
\begin{equation}
<G{(S)}S_i^2>= \Big<G{(S)} \prod_{j \neq i,|j|=1} (S_j^2 cosh(\beta J_{ij} \nabla )+ S_j sinh(\beta J_{ij} \nabla ) + 1 - S_j^2 ) \Big> g(x)|_{x=0}. \label{eq15}
\end{equation}
The function $G(S)$ is any function of S, except $S_i^2$.
The equations (\ref{eq12}) and (\ref{eq15}) are exact and generalize Callen's identity which was obtained for the $S=1/2$ Ising model \cite{Callen1963}.

\section{Exact correlation identities applied to the d=2 and d=3 lattices}
Let us apply the previous results for $<F{(S)}S_i>$ and $<G{(S)}S_i^2>$ given by equations (\ref{eq12}) and (\ref{eq15})
for specific lattices in two- and three-dimensions.The two spins correlation functions, $<S_0 S_l>$, are obtained from equations (\ref{eq12}) and (\ref{eq15}) by defining $F{(S)} = S_l$.

\subsection{For the $d=2$ and $z=3$, the honeycomb lattice}\label{d2z3}
We obtain from Eq.(\ref{eq12})
\begin{eqnarray}
<S_0 S_l> &=& A_1 \sum_i<S_i S_l> + A_2 \sum_{i<j}<S_i S_j^2 S_l> \nonumber \\
&&+ A_3 \sum_{i<j<k}<S_i S_j S_k S_l> + A_4 \sum_{i<j<k}<S_i S_j^2 S_k^2 S_l>, \label{eq16}
\end{eqnarray}
where the A coefficients are given in appendix \ref{Az3}.
We also obtain, from Eq.(\ref{eq15}),
\begin{eqnarray}
<S_0^2 S_l> &=& B_0 + B_1\sum_i <S_i^2 S_l> + B_2 \sum_{i<j}<S_i S_j S_l>  \nonumber \\
&& + B_3 \sum_{i<j}<S_i^2 S_j^2 S_l> + B_4  \sum_{i<j<k}<S_i S_j S_k^2 S_l>  \nonumber \\
&& + B_5  \sum_{i<j<k}<S_i^2 S_j^2 S_k^2 S_l> , \label{eq17}
\end{eqnarray}
where the B coefficients are given in appendix \ref{Az3}.

\subsection{For the $d=2$ and $z=4$, the square lattice}\label{d2z4}
We obtain from Eq.(\ref{eq12}) for the two spin correlation functions $<S_0 S_l>$ the expression,
\begin{eqnarray}
<S_0 S_l> &=& A_1 \sum_i<S_i S_l> + A_2\sum_{i<j} <S_i S_j^2S_l> + A_3\sum_{i<j<k} <S_i S_j S_k S_l> \nonumber\\
&& + A_4\sum_{i<j<k} <S_i S_j^2 S_k^2 S_l> + A_5\sum_{i<j<k<m}<S_i S_j S_k S_m^2 S_l> \nonumber\\
&&+ A_6 \sum_{i<j<k<m}<S_i S_j^2 S_k^2 S_m^2 S_l>,
\end{eqnarray}
where the A coefficients are given in appendix \ref{Az4}.
We also obtain, for the function $<S_0^2 S_l>$,
\begin{eqnarray}
 <S_0^2 S_l> &=& B_0 + B_1  \sum_i<S_i^2 S_l> + B_2\sum_{i<j} <S_i S_j S_l> + B_3 \sum_{i<j}<S_i^2 S_j^2 S_l> \nonumber\\
&&+ B_4 \sum_{i<j<k}<S_i S_j S_k^2 S_l> + B_5 \sum_{i<j<k}<S_i^2 S_j^2 S_k^2 S_l> \nonumber \\
&& + B_6 \sum_{i<j<k<m}<S_i S_i S_k S_m>  + B_7\sum_{i<j<k<m} <S_i S_j S_k^2 S_m^2> \nonumber \\
&&+ B_8 \sum_{i<j<k<m}<S_i^2 S_j^2 S_k^2 S_m^2 S_l>, \label{eq19}
\end{eqnarray}
where the B coefficients are given in appendix \ref{Az4}.

\subsection{For the $d=3$ and $z=6$, the cubic lattice}\label{d3z6}
We obtain from Eq.(\ref{eq12})
\begin{eqnarray}
<S_0 S_l>&=& A_1 \sum_i<S_i S_l> + A_2\sum_{i<j} <S_i S_j^2 S_l> + A_3\sum_{i<j<k} <S_i S_j S_k S_l>\nonumber\\
&&+ A_4\sum_{i<j<k} <S_i S_j^2 S_k^2 S_l>  + A_5\sum_{i<j<k<m}<S_i S_j S_k S_m^2 S_l> \nonumber \\
&&+ A_6 \sum_{i<j<k<m}<S_i S_j^2 S_k^2 S_m^2 S_l> + A_7 \sum_{i<j<k<m<n}<S_i S_j S_k S_m S_n S_l>\nonumber \\
&&+ A_8 \sum_{i<j<k<m<n}<S_i S_j S_k S_m^2 S_n^2 S_l>\nonumber \\
&& + A_9 \sum_{i<j<k<m<n<p}<S_i S_j S_k S_m S_n S_p^2 S_l>\nonumber \\
&&+ A_{10} \sum_{i<j<k<m<n}<S_i S_j^2 S_k^2 S_m^2 S_n^2 S_l> \nonumber \\
&&+ A_{11} \sum_{i<j<k<m<n<p}<S_i S_j S_k S_m^2 S_n^2 S_p^2 S_l>,\label{eq20}
\end{eqnarray}
where the A coefficients are given in appendix \ref{Az6}.
We also obtain, for the function $<S_0^2 S_l>$,
\begin{eqnarray}
<S_0^2 S_l> &=& B_0 + B_1  \sum_i<S_i^2 S_l> + B_2\sum_{i<j} <S_i S_j S_l> + B_3 \sum_{i<j}<S_i^2 S_j^2 S_l> \nonumber\\
&&+ B_4 \sum_{i<j<k}<S_i S_j S_k^2 S_l> + B_5 \sum_{i<j<k}<S_i^2 S_j^2 S_k^2 S_l> \nonumber \\
&& + B_6 \sum_{i<j<k<m}<S_i S_i S_k S_m>  + B_7\sum_{i<j<k<m} <S_i S_j S_k^2 S_m^2> \nonumber \\
&&+ B_8 \sum_{i<j<k<m}<S_i^2 S_j^2 S_k^2 S_m^2 S_l>  \nonumber\\
&&+ B_9 \sum_{i<j<k<m<n<p} <S_i S_j S_k S_m S_n S_p S_l> \nonumber\\
&&+B_{10} \sum_{i<j<k<m<n} <S_i^2 S_j^2 S_k^2 S_m^2 S_n^2 S_l>\nonumber\\
&& +B_{11} \sum_{i<j<k<m<n<p} <S_i^2 S_j^2 S_k^2 S_m^2 S_n^2 S_p^2 S_l> \nonumber\\
&&+ B_{12} \sum_{i<j<k<m<n} <S_i^2 S_j^2 S_k^2 S_m S_n S_l>\nonumber\\
&& + B_{13} \sum_{i<j<k<m<n<p} <S_i^2 S_j^2 S_k^2 S_m^2 S_n S_p S_l> \nonumber\\
&&+ B_{14} \sum_{i<j<k<m<n} <S_i^2 S_j S_k S_m S_n S_l>\nonumber\\
&&+ B_{15} \sum_{i<j<k<m<n<p} <S_i^2 S_j^2 S_k S_m S_n S_p S_l>, 
\end{eqnarray}
where the B coefficients are given in appendix \ref{Az6}.

The sums over $i, j, k, m, n$ and $p$ are over the nearest neighbors of $0$ to which we have given a numerical ordering. The proof of results (\ref{eq16}) for the case (\ref{d2z3}), the honeycomb lattice, is presented in the appendix \ref{Aproof}, as an example for the other cases.

\section{Application of the correlation inequalities}

In the following results we will made use of the following inequalities: $<S_A> \geq 0$ (Griffiths I),  $<S_A S_B> - <S_A><S_B>  \geq 0$ (Griffiths II) (see \cite{Griffiths1969}, \cite{Sylvester1976},\cite{Fernandez},\cite{Szasz1978}), $ <S_i F>  \leq \sum_j <S_i S_j><dF/dS_j>$ (Newman's) (\cite{Newman1975, Sylvester1976})and $<S_i^2 S_A>  \leq  <S_A>$ (\cite{Braga1},\cite{Braga2}), where $<S_A>=\prod_iS_i$, $<S_B>=\prod_iS_i$ and $F$ is a polynomial function of variables $S$.

From the equations for the two spin correlation functions obtained in subsections (\ref{d2z3}) , (\ref{d2z4}) and (\ref{d3z6}) and applying the Griffith's and Newman's inequalities we obtain an inequality of the form

\begin{equation}
<S_0 S_l> \leq \sum_{ |i|=1} a_i <S_i S_l>,
\end{equation}
 where $a_i$ is a sum of products of two-point functions.
\\
\\
(a) Case $d=2$, $z=3$, honeycomb lattice.
\\
\\
Using 
\begin{equation}
<S_j^2S_iS_l> \leq <S_iS_l> \label{eq22}
\end{equation}
in equation (\ref{eq19}), in the $A_2$ term and Griffiths II , i.e., 
\begin{equation}
<S_iS_jS_kS_l> \geq <S_iS_j><S_kS_l>
\end{equation}
 in the $A_3$ term, and noticing that $A_2$ and $A_3$ are negative, we get for d=2, z=3,
\begin{equation}
<S_0 S_l> \leq (A_1-\mid A_2\mid -\mid A_3\mid <S_0 S_1>_{1D} + A_4) \sum _{\mid i=1\mid} <S_i S_l>,
\end{equation}
\\
\\
(b) Case $d=2$, $z=4$, square lattice.
\\
\\
Using inequality (\ref{eq22}) in equation (\ref{eq20}), in the $A_2$ term $(A_2<0)$ term, Griffiths II in the $A_3$ term $(A_3<0)$, the inequalities
\begin{equation}
<S_j^2S_k^2S_iS_l> \leq <S_iS_l> \label{eq26}
\end{equation} 
in the $A_4$ term $(A_4>0)$ and
\begin{equation}
<S_j^2S_k^2S_m^2S_iS_l> \leq <S_iS_l>
\end{equation} 
in the $A_6$ term $(A_6>0)$ and in the $A_5$ term using  Griffiths II, we get for d=2, z=4,
\begin{eqnarray}
<S_0 S_l> &\leq& (A_1 - \mid A_2\mid - <S_1 S_2>_{1D} \mid A_3\mid   \nonumber \\
&& + A_4+ <S_1 S_2>_{1D} A_5 + A_6) \sum _{\mid i=1\mid}<S_i S_l>,
\end{eqnarray}
\\
\\
(c) Case $d=3$, $z=6$, cubic lattice.
\\
\\
As before, we use in equation (\ref{eq20}), inequality (\ref{eq22}) in the $A_2$ term $(A_2<0)$ term, Griffiths II in the $A_3$ term $(A_3<0)$, the inequalities (\ref{eq26}) in the $A_4$ term $(A_4>0)$, inequality (24) in the $A_6$ term $(A_6>0)$ , and Griffiths II in the $A_5$ term. For the term $A_7(>0)$ we use Newman's inequality and for the terms $A_8(>0)$, $A_9(>0)$, $A_{10}(>0)$ and $A_{11}(>0)$, we use inequality (22). Then, we get for d=3, z=6,
\begin{eqnarray}
<S_0 S_l> &\leq& (A_1 - \mid A_2\mid - <S_1 S_2>_{1D} \mid A_3\mid  + A_4 \nonumber \\
&& + <S_1 S_2>_{1D} A_5 \nonumber\\
&&+ A_6 +A_7+A_8+A_9+A_{10}+A_{11}) \sum _{\mid i=1\mid}<S_i S_l>
\end{eqnarray}

The two-spin correlation function $<S_1 S_2>_{1D}$ is the one-dimension model two spin correlation separated by a distance of two lattice sites.
By bounding the resulting two-point function occurring in the previous results from below with the two-point function of the one-dimensional infinite chain (see Appendix \ref{Aproof}), we get: 

\begin{equation}
<S_0 S_l> \leq \sum_{\mid i=1\mid} a_i <S_i S_l>,
\end{equation}

where,
\\
\\(a) For $d=2$, $z=3$, honeycomb lattice.
\begin{equation}
a_j = A_1-\mid A_2\mid -\mid A_3\mid <S_0 S_1>_{1D} + A_4 ;
\end{equation} \\
(b)For $d=2$, $z=4$, square lattice.
\begin{eqnarray}
&&a_j= A_1 - \mid A_2\mid - <S_1 S_2>_{1D} \mid A_3\mid  + A_4\nonumber\\
&&+ <S_1 S_2>_{1D} A_5 + A_6;
\end{eqnarray}\\
(c)For $d=3$, $z=6$, cubic lattice.
\begin{eqnarray}
a_j &=& A_1 - \mid A_2\mid - <S_1 S_2>_{1D} \mid A_3\mid  + A_4 \nonumber \\
&& + <S_1 S_2>_{1D} A_5 + A_6 \nonumber\\
&&+A_7+A_8+A_9+A_{10}+A_{11}.
\end{eqnarray}

The one-dimensional correlation function is given by (see Appendix \ref{Ac1d}):
\begin{equation} 
<S_1 S_2>_{1D} = \frac{1+\sqrt{(1-2f(2\beta J)}}{f(2\beta J)}
\end{equation}
and $f(2\beta J)$ is given by (\ref{eq10}).

\section{Numerical results}

Evaluating numerically the value of $T$ such that $\sum a_j \leq  1$, $a_j > 0$, we obtain, by sufficient condition, upper bounds for $T_c$, which are shown in tables \ref{tab1} and \ref{tab2}, in comparison with results obtained by other methods. 

\begin{table}
\centering
\caption{Estimatives for $k T_C/J$ for $D=0$ in previous and in the present work.} 
 
\begin{tabular}{cccc} 

\hline
  &   $d=2, z=3$   &   $d=2, z=4$   &   $d=3, z=6$   \\
\hline
MFA & $2$ & $2.667$ & $4$ \\ 
Siqueira & $1.518$ & $2.188$ & $3.516$ \\
Yuksel & - & $1.964$ & - \\
CVM & - & - & $2.886$ \\
Series & - & $1.688$ & $3.192$ \\
RG & - & $2.128$ & $3.474$ \\
Monte Carlo \cite{Beale1986} & - & $1.695$ & - \\
Monte Carlo \cite{Plascak1998} & - & $1.681$ & - \\
Wang Landau \cite{Plascak2006} & - & $1.714$ & - \\
Present work & $1.591$ & $2.322$ & $3.678$ \\
\hline
\end{tabular}
\label{tab1}
\end{table}

\begin{table}
\centering
\caption{Estimatives for $k T_C/J$ for $D=\infty$ in previous and in the present work.} 
 
\begin{tabular}{cccc} 

\hline
  &   $d=2, z=3$   &   $d=2, z=4$   &   $d=3, z=6$   \\
\hline
MFA & $3$ & $4$ & $6$ \\ 
Siqueira & $2.103$ & $3.088$ & $5.076$ \\
CVM & - & - & $3.876$ \\
Series & - & - & $4.482$ \\
RG & - & $2.884$ & $4.932$ \\
Present work & $1.999$ & $3.070$ & $5.084$ \\
\hline
\end{tabular}
\label{tab2}
\end{table}

For the evaluation of the self-correlation terms ($<S_i^2>$) that emerge from the application of the Griffith's and Newman's inequalities, we use, for the $D=0$ case, $<S_i^2> \leq 2/3$, correct for a spin 1 ferromagnetic system, and, for the $D = \infty$ case, $<S_i^2> = 1$, since in this limit the $S_i=0$ spin value is suppressed.

For the honeycomb lattice our result has to be compared with the mean field and the effective field calculations. Those results are not rigorous, as ours, and the  numerical values we obtain improve those mean field type results and therefore represent the upper bounds, for the limits $D=0$ and $D=\infty$. For the square and cubic lattices besides the mean field type results, for which the previous comments apply, there are other results, better than mean field type, obtained by series and renormalization group calculations, which can be used as a comparison. The importance of the present numerical results lies in the fact that they were obtained using an identity and rigorous inequalities for the two-spin correlation function. For this reason they represent rigorous upper bounds for the critical temperature. 

\section{Final Comments}
We have presented the derivation of correlation identities for the Blume-Capel spin-1 model which are exact in all dimensions, and we have made use of correlation inequalities to obtain the upper bounds for the transition temperature. The coupling constants obtained for those bounds are calculated for d=2 (honeycomb and square lattices) and d=3 (cubic lattice). We obtain rigorous results that improve mean field type calculations.

\appendix

\section{Coefficients of the Spin Correlation Identities for d=2, z=3 and z=4.}

With $k=\beta J$ and $f(x)$ given by relation (\ref{eq10}), we have for

\subsection{d=2, z=3}\label{Az3}

\begin{equation}
A_1 = 3 f(k) >0,
\end{equation}
\begin{equation}
A_2 = (3f(2k) - 6f(k))<0,
\end{equation}
\begin{equation}
A_3 = \frac{1}{4} \big( f(3k) - 3f(k)  \big) <0
\end{equation}
\begin{equation}
A_4 = \frac{3}{4} \big( 5f(k) + f(3k) - 4 f(2k) \big) >0
\end{equation}
and
\begin{equation}
B_0 = g(0),
\end{equation}
\begin{equation}
B_1 = 3 (g(k)-g(0)),
\end{equation}
\begin{equation}
B_2 = \frac{3}{2} (g(2k) - g(0)),
\end{equation}
\begin{equation}
B_3 = \frac{3}{2}g(2k) + - 6 g(k) + \frac{9}{2}g(0),
\end{equation}
\begin{equation}
B_4 = \frac{3}{4} ( g(3k) - g(k) -2g(2k) + 2g(0)),
\end{equation}
\begin{equation}
B_5 = \frac{1}{4}g(3k) -\frac{3}{2}g(2k) + \frac{15}{4}g(k) - \frac{5}{2}g(0).
\end{equation}

\subsection{d=2, z=4}\label{Az4}
\begin{equation}
A_1=4 f(k) >0,
\end{equation}
\begin{equation}
A_2= 6 f(2k) - 12 f(k) <0,
\end{equation}
\begin{equation}
A_3= f(3k)-3f(k) <0,
\end{equation}
\begin{equation}
A_4= 15 f(k) -  12 f(2k) +  3 f(3k))>0,
\end{equation}
\begin{equation}
A_5= \frac{1}{2} f(4k) - f(3k) -f(2k) +3f(k)>0
\end{equation}
\begin{equation}
A_6 = \frac{1}{2} f(4k) -3 f(3k) + 7 f(2k) - 7 f(k) < 0
\end{equation}
and
\begin{equation}
B_0 = g(0),
\end{equation}
\begin{equation}
B_1 = 4(g(k)-g(0)),
\end{equation}
\begin{equation}
B_2 = 3 (g(2k)-g(0)),
\end{equation}
\begin{equation}
B_3 = 3 (g(2k) - 4g(k) + 3 g(0)),
\end{equation}
\begin{equation}
B_4 = 3 (g(3k) -2g(2k) -g(k) +2g(0)),
\end{equation}
\begin{equation}
B_5 = g(3k) -6g(2k)+15g(k)-10g(0),
\end{equation}
\begin{equation}
B_6 = \frac{1}{8} (g(4k) -4g(2k) + 3g(0)),
\end{equation}
\begin{equation}
B_7 = \frac{3}{4}g(4k)-3g(3k)+3g(2k)+3g(k)-\frac{15}{4}g(0),
\end{equation}
\begin{equation}
B_8 = \frac{1}{8}g(4k)-g(3k)+\frac{7}{2}g(2k)-7g(k)+\frac{35}{8}g(0).
\end{equation}

\subsection{d=3, z=6}\label{Az6}
\begin{equation}
A_1=6f(k)>0,
\end{equation}
\begin{equation}
A_2=-30f(k)+15f(2k)<0,
\end{equation}
\begin{equation}
A_3=5f(3k)-15f(k)<0,
\end{equation}
\begin{equation}
A_4=75f(k)+15f(3k)-60f(2k)>0,
\end{equation}
\begin{equation}
A_5=-15f(3k)+45f(k)+\frac{15}{2}f(4k)-15f(2k)>0,
\end{equation}
\begin{equation}
A_6=-45f(3k)-105 (f(k) - f(2k))+\frac{15}{2}f(4k)<0,
\end{equation}
\begin{equation}
A_7=\frac{3}{8}f(5k)-\frac{15}{8}f(3k)+\frac{15}{4}f(k)>0,
\end{equation}
\begin{equation}
A_8=\frac{45}{4}f(3k)-\frac{105}{2}f(k)+\frac{15}{4}f(5k)-15f(4k)+30f(2k)<0,
\end{equation}
\begin{equation}
A_9=-\frac{3}{8}f(5k)+\frac{15}{8}f(3k)-\frac{15}{4}f(k)+\frac{3}{16}f(6k)-\frac{3}{4}f(4k)+\frac{15}{16}f(2k)<0,
\end{equation}
\begin{equation}
A_{10}= \frac{405}{8}f(3k)+\frac{315}{4}f(k)+\frac{15}{8}f(5k)-15f(4k)-90f(2k)>0,
\end{equation}
\begin{equation}
A_{11}=-\frac{5}{4}f(3k)+\frac{45}{2}f(k)+\frac{15}{2}f(4k)-\frac{135}{8}f(2k)-\frac{15}{4}f(5k)+\frac{5}{8}f(6k)>0
\end{equation}
and
\begin{equation}
B_0 = g(0),
\end{equation}
\begin{equation}
B_1 = 6(g(k)-g(0)),
\end{equation}
\begin{equation}
B_2 = \frac{15}{2} (g(2k)-g(0)),
\end{equation}
\begin{equation}
B_3 = \frac{15}{2} (g(2k) - 4g(k) + 3 g(0)),
\end{equation}
\begin{equation}
B_4 = 15 (g(3k) -2g(2k) -g(k) +2g(0)),
\end{equation}
\begin{equation}
B_5 = 5 (g(3k) -6g(2k)+15g(k)-10g(0)),
\end{equation}
\begin{equation}
B_6 = \frac{15}{8} (g(4k) -4g(2k) + 3g(0)),
\end{equation}
\begin{equation}
B_7 = 45 \big( \frac{1}{4}g(4k) - g(3k) + g(2k) + g(k) -\frac{5}{4} g(0) \big),
\end{equation}
\begin{equation}
B_8 = 15 \big( \frac{1}{8}g(4k)-g(3k)+\frac{7}{2}g(2k)-7g(k)+\frac{35}{8}g(0) \big),
\end{equation}
\begin{equation}
B_9 = \frac{1}{32} (g(6k)-6g(4k)+15g(2k)-10g(0)),
\end{equation}
\begin{equation}
B_{10} = \frac{3}{8} (-126g(0)+45g(3k)+210g(k)-120g(2k)-10g(4k)+g(5k)),
\end{equation}
\begin{equation}
B_{11} = \frac{3}{8} \big( -\frac{55}{3}g(3k)-66g(k)+\frac{165}{4}g(2k)+\frac{77}{2}g(0)+\frac{1}{12}g(6k)+\frac{11}{2}g(4k)-g(5k) \big),
\end{equation}
\begin{equation}
B_{12} = \frac{15}{4} (-8g(2k)+14g(0)+g(5k)-14g(k)+13g(3k)-6g(4k)) ,
\end{equation}
\begin{equation}
B_{13} = \frac{15}{32} (-40g(3k)+48g(k)+15g(2k)-42g(0)+26g(4k)+g(6k)-8g(5k)) ,
\end{equation}
\begin{equation}
B_{14} = \frac{15}{8} (-2g(4k)+8g(2k)-6g(0)+g(5k)-3g(3k)+2g(k)) ,
\end{equation}
\begin{equation}
B_{15} = \frac{15}{32} (2g(4k)-17g(2k)+14g(0)-4g(5k)+12g(3k)-8g(k)+g(6k)).
\end{equation}

\section{Proof of the correlation identity  for the honeycomb lattice}\label{Aproof}

From equation (\ref{eq12})
\begin{equation}
<F{(S)} S_i>= \Big< F{(S)} \prod_{j \neq i} (S_j^2 cosh(\beta J_{ij} \nabla ) + S_j sinh(\beta J_{ij} \nabla ) + 1 - S_j^2 ) \Big> f(x)|_{x=0} \label{eq70}
\end{equation}
where,
\begin{equation}
f(x)=\frac{2 e^{\beta D} sinh(x)}{2 e^{\beta D} cosh(x) + 1}, \label{eq71}
\end{equation}
we obtain $<S_0 S_l>$, for the honeycomb lattice,
\begin{eqnarray}
&&<S_0 S_l> = <S_l ( 1+ S_1 \sinh J\nabla + S_1^2 [\cosh J\nabla - 1]) \nonumber \\
&&\times ( 1+ S_2 \sinh J\nabla + S_2^2 [\cosh J\nabla - 1]) \nonumber \\
&&\times ( 1+ S_3 \sinh J\nabla + S_3^2 [\cosh J\nabla - 1])> \nonumber \\
\end{eqnarray}
 where $S_1$, $S_2$ and $S_3$ are the neighbours of $S_0$.

Or,
\begin{eqnarray}
&&<S_0 S_l> = 3 a_1<S_1 S_l> + 6(a_2 - a_1)<S_1 S_2^2> \nonumber \\
&&+ a_3 <S_1S_2S_3>+(a_1 - 2 a_2 + a_4) <S_1 S_2^2 S_3^2>,
\end{eqnarray}
where,
\begin{eqnarray}
&&a_1 = \sinh J\nabla \cdot f(x)\mid_{x=0}  = f(\beta J)\nonumber \\ 
&&a_2 = \sinh J \nabla \cosh J \nabla  \cdot f(x)\mid_{x=0} = 1/2 f(2\beta J)\nonumber \\
&&a_3 = \sinh J^3{\nabla} \cdot f(x)\mid_{x=0} = 1/4[f(3\beta J) - 3f(\beta J)]\nonumber \\
&&a_4 = \sinh J\nabla \cosh^2 J\nabla  \cdot f(x)\mid_{x=0} = 1/4[ f(3\beta J)+f(\beta J)]
\end{eqnarray}

From those results we obtain equations (\ref{eq16}) and (\ref{eq17}) of section \ref{d2z3}. 

\section{Spin Correlation for the One-Dimensional S=1 Blume-Capel Model}\label{Ac1d}

For the linear chain, we have,
\begin{eqnarray}
&&<S_0> = <( 1+ S_1 \sinh J\nabla + S_1^2 [\cosh J\nabla - 1]) \nonumber \\
&& ( 1+ S_{-1} \sinh J\nabla + S_{-1}^2 [\cosh J\nabla - 1])> \cdot f(x)\mid _{x=0} \nonumber \\
\end{eqnarray}
with $f(x)$ given by expression (\ref{eq10}) and $S_1$ and $S_{-1}$ are neighbors of $S_0$.
We obtain for the two-spin correlation function
\begin{eqnarray}
&&<S_0 S_R> = <(S_1 S_R + S_{-1}S_R)> f(k)\nonumber \\ 
&&+ <(S_1S_{-1}S_1S_R + S_1S_{-1}S_{-1}S_R)> (\frac{1}{2}f(2k)-f(k))
\end{eqnarray}
where $k = \beta J$.
Applying the inequalities (\cite{Braga1},\cite{Braga2})
\begin{eqnarray}
&&<S_1^2S_{-1}S_R>   \leq   <S_{-1}S_R> \nonumber \\ 
&&<S_{-1}^2S_1S_R>   \leq   <S_1S_R> 
\end{eqnarray}
we get,
\begin{eqnarray}
&&<S_0 S_R>  \leq  (<S_1 S_R> + <S_{-1}S_R>) f(k)\nonumber \\ 
&&+ (<S_{-1}S_R> + <S_1S_R>) [1/2f(2k)-f(k)]
\end{eqnarray}
Defining $C(R) = <S_0S_R>$ we get
\begin{equation}
C(R) = A(k) (C(R+1) + C(R-1)), 
\end{equation}
where $A(k)= f(2k)/2$.

If  $\gamma (R) = C(R+1)/C(R)$ is inserted in the previous equation we get 
\begin{equation}
1 = A(k)(\gamma(R) + \gamma(R)^{-1}).
\end{equation}

So, $C(R) =  \gamma ^R$ and 
\begin{equation}
\gamma = \frac{1+\sqrt{1-2f(2\beta J)}}{f(2\beta J)}. 
\end{equation}

\section*{Acknowledgements} FCSB is grateful to CAPES/Brazil for the financial support that made possible his visit to the UFSJ/Brazil. ALM acknowledges financial support from CNPq-Brazil and FAPEMIG-Brazil.
\vspace{1cm}
 

\vspace{1cm}

\end{document}